\documentclass[twocolumn,aps,prl,10pt,amsmath,amssymb,nofootinbib,showpacs,superscriptaddress,floatfix]{revtex4-1}

\DeclareFontFamily{U}{rcjhbltx}{}
\DeclareFontShape{U}{rcjhbltx}{m}{n}{<->rcjhbltx}{}
\DeclareSymbolFont{hebrewletters}{U}{rcjhbltx}{m}{n}

\usepackage{graphicx}
\usepackage{color}
\usepackage[usenames,dvipsnames]{xcolor}
\usepackage[colorlinks=true,linkcolor=Red,citecolor=Green,linktoc=page]{hyperref}
\usepackage{multirow}
\usepackage{float}
\usepackage{flushend}
\usepackage{balance}
\usepackage[varg]{txfonts}
\usepackage{ulem}
\usepackage{fancyhdr}

\DeclareMathSymbol{\lamed}{\mathord}{hebrewletters}{108}

\begin{document}
\title{Emergent time, cosmological constant and boundary dimension at infinity \\ in combinatorial quantum gravity}



\author{C.\,A.\,Trugenberger}

\affiliation{SwissScientific Technologies SA, rue du Rhone 59, CH-1204 Geneva, Switzerland}



\begin{abstract}
Combinatorial quantum gravity is governed by a discrete Einstein-Hilbert action formulated on an ensemble of random graphs. There is strong evidence for a second-order quantum phase transition separating a random phase at strong coupling from an ordered, geometric phase at weak coupling. Here we derive the picture of space-time that emerges in the geometric phase, given such a continuous phase transition. In the geometric phase, ground-state graphs are discretizations of Riemannian, negative-curvature Cartan-Hadamard manifolds. On such manifolds, diffusion is ballistic. Asymptotically, diffusion time is soldered with a manifold coordinate and, consequently,
the probability distribution is governed by the wave equation on the corresponding Lorentzian manifold of positive curvature, de Sitter space-time. With this asymptotic Lorentzian picture, the original negative curvature of the Riemannian manifold turns into a positive cosmological constant. The Lorentzian picture, however, is valid only asymptotically and cannot be extrapolated back in coordinate time. Before a certain epoch, coordinate time looses its meaning and the universe is a negative-curvature Riemannian ``shuttlecock" with ballistic diffusion, thereby avoiding a big bang singularity. The emerging coordinate time leads to a de Sitter version of the holographic principle relating the bulk isometries with boundary conformal transformations. While the topological boundary dimension is $(D-1)$, the so-called ``dimension at infinity" of negative curvature manifolds, i.e. the large-scale spectral dimension seen by diffusion processes with no spectral gap, those that can probe the geometry at infinity, is always three. 
\end{abstract}
\maketitle

\section{Introduction}

General relativity (GR), as a quantum field theory of space-time, is perturbatively non-renormalizable. Decreasing the scale of observation requires more and more parameters to compensate the proliferating divergences so that the model looses all predictive power at high energies. Various avenues have been proposed to cure this problem. One is to consider GR as an effective field theory, i.e. to abandon the tenets of local quantum field theory below the Planck scale in favour of a different principle, like string theory (see e.g. \cite{strings}) or causal set theory (for a review see \cite{causalsets}). Another approach is to look for symmetry relations that reduce the number of required parameters to a finite set of physical couplings, the asymptotic safety programme (for a review see \cite{as}). Finally, one can try a non-perturbative quantization approach, as in loop quantum gravity (for a review see \cite{loopqg}), or by formulating quantum gravity on a ``lattice". This latter procedure is based on discretizing space (or space-time) in terms of simplices, generalized triangles, and is correspondingly called dynamical triangulation (DT) in its original Euclidean formulation, or causal dynamical triangulation (CDT) in its Lorentzian form (for a review see \cite{cdt1,cdt2}). Another discrete approach is the tensor model, again either in its original formulation \cite{tensor1, tensor2} or in its recent canonical version with an incorporated temporal dimension \cite{sasakura}. 

Common to all these approaches is the role of time and causality, built in the models from the very beginning. In causal set theory, causality is the very defining principle of the construction. But even the discrete approach works only in its Lorentzian formulation (CDTs), in which the simplices are organized to define a foliation of the manifold they discretize into space and time, \cite{cdt1, cdt2}. As Hartle and Hawking showed in their seminal contribution \cite{hh} (for a review see \cite{hhrev}), however there are strong indications that the quantum mechanical evolution of the universe is dominated by one ``history" in which the universe initially had an Euclidean metric and that time as we know it is far from fundamental but most probably an emergent concept. This point of view is shared by Polyakov, who suggests that the very existence of antiparticles is a hint of an underlying Euclidean structure of space-time \cite{polyakov}.

Triangulations by simplices still represent piece-wise flat manifolds. As an alternative, we proposed to formulate quantum gravity on much ``wilder" and abstract structures, random graphs (for a review see \cite{graphrev}), using the combinatorial Ollivier-Ricci curvature \cite{olli0, olli1, olli2} (see also \cite{olli3, olli4}) to construct a graph-theoretic Einstein-Hilbert action \cite{comb1}. The Ollivier curvature is only one of several possible notions of curvature on random graphs. However, it is the only one that has been shown to converge to continuum Ricci curvature for geometric random graphs defined on manifolds \cite{conv1, conv2} (see \cite{tee} for its relation to Forman curvature \cite{forman}). It is computationally cumbersome, which prompted the definition of more tractable approximations \cite{klit1, klit2, klit3}, but it simplifies substantially on a certain class of ``physical" random graphs which are those relevant for the quantum gravity model \cite{comb1, comb2, gorsky}. 

In this quantum gravity model there is no a priori notion of either time or space: the model is defined on random combinatorial structures and it was correspondingly named {\it combinatorial quantum gravity}. There is, however, strong numerical evidence that geometric manifolds self-assemble from the random graphs in a second-order phase transition corresponding to the condensation of 4-cycles, squares on the network \cite{comb1, comb2}, in concert with previous observations that the emergence of geometry from random graphs is tied to clustering \cite{dall, krioukov}. This was explicitly confirmed for cubic graphs, an essentially exactly solvable version of the model, which were shown to self-assemble into 1D manifolds \cite{comb3}. The role played by the Ollivier curvature for the geometric properties of general graph models has also been recently discussed in \cite{gosztolai}. 

The 1D case is simple because there is no geometry, only topology. The question arises, though, what manifolds emerge in higher dimensions and how does time come into play. Here we focus on the consequences of the second-order phase transition, assuming that its nature can be further cemented by large-scale numerical simulations presently out of reach of the available computational resources. In particular, we show that the picture of space-time emerging from combinatorial quantum gravity is reminiscent of the Hartle-Hawking scenario \cite{hh, hhrev}. Below a running Planck scale, space decomposes into random graphs. Above this scale, graphs assemble into Riemannian, negative-curvature Cartan-Hadamard (CH) manifolds (see e.g. \cite{shiga}). We shall provide strong numerical evidence to support this analytical picture in 2D. In higher dimensions, extracting the continuum properties from sufficiently large graphs is beyond the presently available computational power, although the phase transition is still clearly identified. We expect that resources at high-performance computing centres may be sufficient to address at least the 3D case. 

The negative curvature of the ground state was recently identified also using the above mentioned \cite{klit1, klit2} Ollivier curvature approximation in 2D \cite{lollqf}. It was, however dismissed in favour of a new concept of ``quantum flatness" because of the lack of a mechanism for the spontaneous emergence of a curvature scale. In combinatorial quantum gravity, on the contrary,  such a curvature scale emerges indeed from the typical correlation length characterizing a second-order phase transition. As we will show, this is the Planck length and is directly related to the cosmological constant and the emergence of time.  

Based on the exact solution in 1D \cite{comb3} and on the strong analytical and numerical arguments in 2D we shall assume that the ground states of combinatorial quantum gravity are negative-curvature CH manifolds in any dimension, including 4D. In the last two sections we shall explore what are the generic consequences.

On negative-curvature CH manifolds diffusion is anomalous \cite{dunne, sandev, procaccia} and, in particular, ballistic \cite{prat, kendall, hsu, hsubrief, arnaudon} (for a review see \cite{hsubook}). Asymptotically for large values of the radial coordinate of the CH hyperboloid model, it is equivalently described by the wave equation on the associated Lorentzian hyperboloid of positive curvature \cite{sandev, chandrashekar}, i.e. de Sitter space-time (for a review see \cite{strominger}).  In this picture, thus, time and Lorentzian metrics are non-fundamental: they are just alternative, approximate descriptions of ballistic diffusion on negative-curvature Riemannian CH manifolds, which become better and better asymptotically. This means that the concept of time as a coordinate of a space-time manifold is an emergent asymptotic concept which cannot be extrapolated backwards: before a certain epoch the universe is a negatively curved Riemannian ``shuttlecock", thereby avoiding big bang singularities, exactly as in the Hartle-Hawking scenario \cite{hh, hhrev}. The existence of a fixed velocity of propagation is traced to the asymptotic properties of diffusion on negatively curved CH manifolds. In the equivalent Lorentzian description, the negative curvature of the fundamental CH manifold becomes the positive cosmological constant of the associated de-Sitter space-time \cite{cosconst}. 

Finally, we shall focus on the large-scale geometry. Simply connected, geodesically complete CH manifolds of (constant) negative curvature have a boundary manifold on which interior isometries are realized as conformal transformations, a Riemannian version of the holographic principle \cite{thooft, susskind} (for a review see \cite{bousso}). Via the emerging time coordinate, however, this can be interpreted as an asymptotic de-Sitter version of holography. While the boundary has topological dimension $(D-1)$ as in the familiar case, it is also characterized by a pseudo-dimension, or dimension at infinity \cite{anker} which describes its spectral dimension as seen by co-diffusing observers, those for which relative diffusion processes have no spectral gap and can thus probe the geometry up to the boundary at infinity, contrary to the usual Brownian motion. This dimension at infinity is always three.

\section{Combinatorial quantum gravity}

Combinatorial quantum gravity \cite{comb1, comb2} is formulated on the configuration space of all 2D-regular graphs on $N$ vertices that satisfy the condition of independent short-cycles. By short cycles we mean those that matter in a discrete version of locality on a graph,  i.e. triangles, squares and pentagons \cite{olli3}. The independent short-cycle condition states that only graphs are admitted for which such cycles do not share more than one edge with each other and can be formulated also as an excluded subgraph condition \cite{comb2}. This condition is a crucial ingredient of the model: it is the loop equivalent of the hard-core condition for bosonic point particles. As the hard-core condition prevents the infinite compressibility of Bose gases, the independent short cycles condition prevents graphs to collapse onto themselves upon loop condensation by requiring that loops can ``touch" but not ``overlap" on more than one edge. It must imposed for any D and will be henceforth implemented in the rest of the paper.

The action is the total Ollivier-Ricci curvature, a discrete graph equivalent of the Einstein-Hilbert action \cite{comb1, comb2}, 
\begin{equation}
S_{\rm EH} = -{2D\over g} \sum_{i \in G} \kappa (i) = -{2D\over g}\sum_{i\in G} \sum_{j\sim i} \kappa (ij) \ ,
\label{deh}
\end{equation}
where we denote by $j \sim i$ the neighbour vertices $j$ to vertex $i$ in graph $G$, i.e. those connected to $i$ by one edge $(ij)$, $\kappa (ij)$ is the coarse Ollivier-Ricci curvature of edge $(ij)$ \cite{olli0, olli1, olli2, olli3, olli4} and $g$ is the coupling constant. The overall factor $2D$ simply defines the overall normalization of the action. 

The Ollivier-Ricci curvature is cumbersome to compute for generic graphs. On $2D$-regular graphs satisfying the independent short cycle condition, however, it becomes very simple \cite{comb2}, 
\begin{equation}
\kappa(ij)=\frac{T_{ij}} {2D}-\left[1-\frac{2+T_{ij} +S_{ij}}{2D}\right]_+ - \left [1-\frac{2+T_{ij} +S_{ij} +P_{ij}}{2D}\right]_+ \ .
\label{orc}
\end{equation}
where the subscript ``+" is defined as $[\alpha]_+ = {\rm max} (0, \alpha)$ and $T_{ij}$, $S_{ij}$ and $P_{ij}$ denote the number of triangles, squares and pentagons supported on egde $(ij)$. 

Combinatorial quantum gravity is the theory defined by the partition function
\begin{equation}
{\cal Z} = \sum_{G \in {\rm CS}} {\rm e}^{-{1\over g\hbar} S_{\rm EH}} \ ,
\label{comqg}
\end{equation} 
where CS denotes the configuration space of $2D$-regular graphs with independent short cycles and $g\hbar$ is the corresponding dimensionless coupling constant.  The model undergoes a quantum phase transition from a strong gravity disordered phase in which the ground state is a locally tree-like random graph to a weak gravity ordered phase in which the ground state contains large numbers of squares but no triangles and pentagons \cite{comb1, comb2}. As usual for network models, phase transitions appear in terms of critical coupling functions of the graph size \cite{graphrev}. In this case we have $\hbar g_{\rm cr} (N) = \hbar \chi_{\rm cr} N^{1-2/D}$, where $\hbar \chi_{\rm cr}$ is the usual numerical critical coupling. This corresponds to a condensation of squares on the graph and in the limit $\hbar g\to 0$ of zero gravity the ground state becomes a flat, locally $D$-dimensional hypercubic lattice with possible residual triangle and pentagon defects. This is what characterizes the parameter $D$ as the dimension of the emerging manifold. In \cite{comb2} strong numerical evidence was provided that this is a second-order quantum phase transition.

\section{Infinite graphs with finite curvature} 

Let us endow the graph edges with a fixed length $\ell$ and let us investigate what are the manifolds that emerge in the geometric phase. 
If we take the na\"ive limit $\hbar g\to 0$ of vanishing gravity, ground state graphs become discretizations of tori ${\rm T}^D$. In the limit $N\to \infty$ the emerging manifolds are the universal coverings of these tori, i.e. Euclidean spaces ${\mathbb R}^D$. This is consistent but not very interesting. Rather, we are interested in the emerging manifolds at weak but finite gravity. In this case the graph curvature will always be negative. 

To proceed we will rely on the typical correlation function associated with second-order phase transitions. In the present case we measure the connected correlation function of the number of squares $S_i$ and $S_j$ at vertices lying at fixed graph distance $d(i,j)=d$,
\begin{equation}
C\left( d \right) = { \langle \left( S_i- \bar S \right) \left( S_j- \bar S \right) \rangle_{d(i,j)=d} \over \langle \left( S- \bar S \right)^2 \rangle} \ ,
\label{correlation}
\end{equation} 
where $\langle \rangle_{d(i,j)=d}$ denotes averages over the graph, in the numerator over all vertex pairs $(i,j)$ at fixed distance $d$, and $\bar S$ is the average number of squares per vertex. For large graph distances this correlation function behaves as
\begin{equation}
C\left( d \right) \propto {\rm e}^{-{d\over \xi}} \ ,
\label{corrlength}
\end{equation}  
with an example of correlation length $\xi (\hbar g) $ shown in Fig.1 for $D=3$ graphs. 

\begin{figure}[t!]
	\includegraphics[width=9cm]{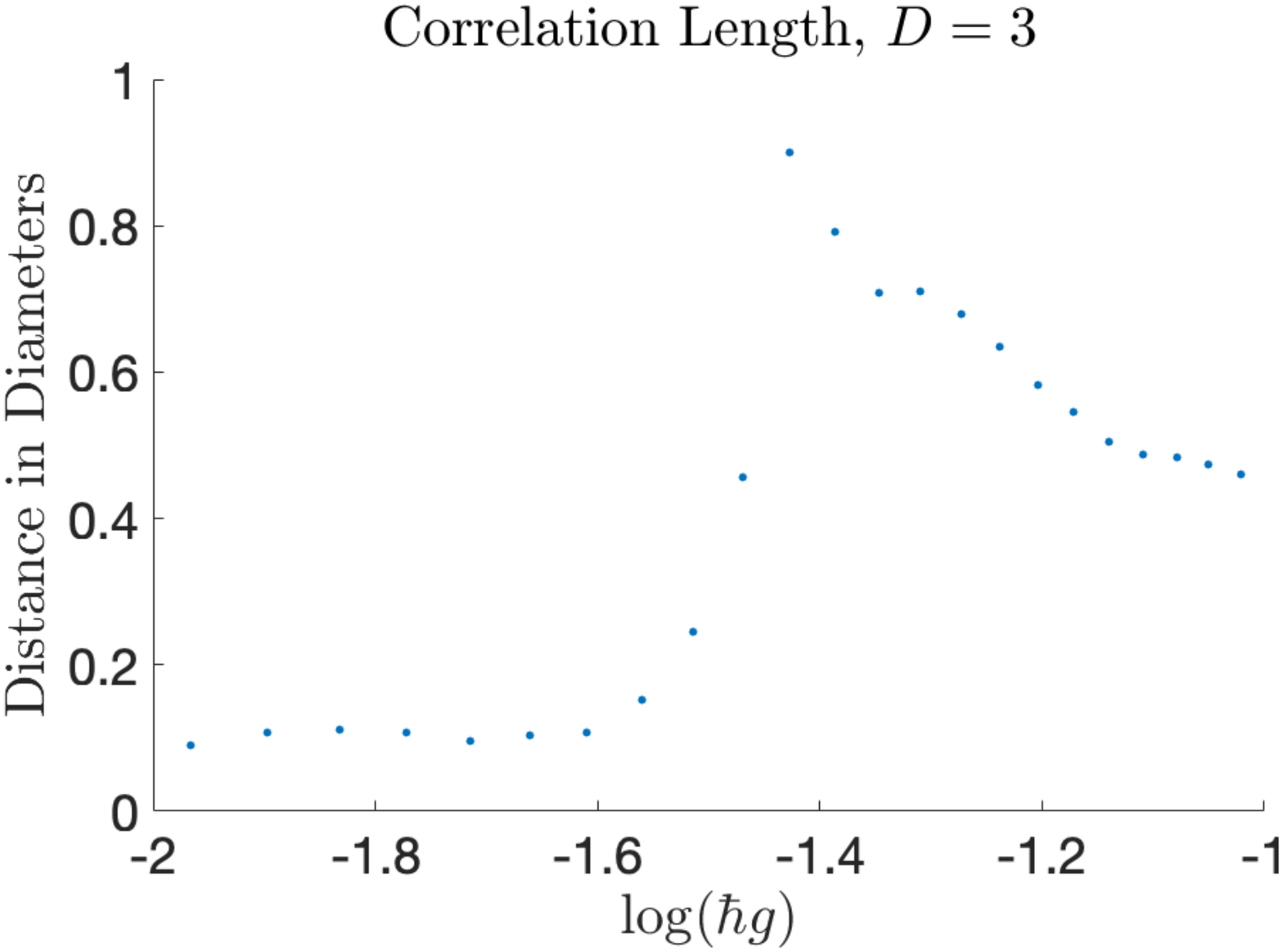}
	\vspace{-0.3cm}
	\caption{The square correlation length as a function of the coupling $\hbar g$ for D=3. }
	\label{Fig.1}
\end{figure}

As expected for a second-order phase transition, the correlation length diverges (up to finite-size effects) at the critical coupling. In the geometric phase, the correlation length is an emergent length scale measuring the distance over which the number of squares, and thus the curvature, vary from the average. 

Let us now consider again the discrete Einstein-Hilbert action 
\begin{equation}
S_{\rm EH} = -{1\over g} \sum_i \kappa_i \ ,
\label{oa1}
\end{equation} 
where the sum runs over the vertices $i$ of the graph, $\kappa_i$ is the Ollivier Ricci scalar at vertex $i$ and we have reabsorbed the normalization factor 2D into $\kappa_i$ for simplicity. As mentioned above, the critical coupling function $g_{\rm cr}(N)$ at the quantum phase transition satisfies $\hbar g_{\rm cr}(N)=N^{1-2/D} \chi_{\rm cr}$, with $\chi_{\rm cr}$ a purely numerical factor \cite{comb1, comb2}.  We can thus rewrite the action as
\begin{equation} 
S_{\rm EH}=-{\hbar \over N^{1-2/D}  \chi_{\rm cr} f} \sum_i \kappa_i \ ,
\label{oa2}
\end{equation}
where we have extracted the $N$-dependence by defining $\hbar g=N^{1-2/D} \chi_{\rm cr} f$, with $f=\chi/\chi_{\rm cr}$ denoting the dimensionless gravitational coupling.

We can label by $z=1\dots N_f = N/(\xi(f)/\ell)^D$ the number of balls of volume $\xi^D(f)$ that fit into the graph at coupling $f$, where $\ell $ is the edge length. By construction, the average curvature $\kappa (f)$ on each of these balls is constant. The emerging manifold is defined by the ensemble of these $N_f$ regions of constant negative curvature $\kappa (f)$. The residual variation of curvature is restricted to the interior of these balls, where the original random graph character can still be detected. We shall identify the dynamical scale $\xi(f)$ with the Planck length $\ell_P(f)$, $({N/N_f})^{1/D} \ell = \ell_P(f)$. We then have  $\ell_P (0) = \ell $  and $\ell_P(1)=N^{1/D} \ell$: at the critical point there are fluctuations in the number of squares on all scales and one needs the whole volume to obtain a ``constant" average $\kappa$. The diverging correlation length at the transition can be used to define a continuum limit and the renormalization of the quantum gravity model, effectively implementing the asymptotic safety scenario. A careful renormalization procedure for a graph model, however, is beyond the scope of the present paper. Here we shall be interested in curvature effects for large graphs.

To this end we shall use the above definitions to rewrite the discrete EH action as, 
\begin{equation}
S_{\rm EH}=-{\hbar \over N^{1-2/D} \ell_P^{D-2}(f) \ \chi_{\rm cr} f} {N\over N_f} \sum_z \ell_P^D {\kappa(f) \over \ell_P^2(f)} \ ,
\label{oa3}
\end{equation}
Identifying $\ell_P^{D-2}(f)/\hbar = G_N(f)$ with the physical Newton constant we get finally,
\begin{equation}
S_{\rm EH}=-{1\over G_N (f) } \sum_z \ell_P^D(f) \ {\kappa(f) N^{2/D} \over f N_f} {1\over \ell_P^2(f)} \ ,
\label{oa4}
\end{equation}
The only quantity here that is not expressed in terms of physical scales is 
\begin{equation}
{\kappa(f)  N^{2/D} \over fN_f} \ ,
\label{renorm}
\end{equation}
where $\kappa(f)$ is a function such that
\begin{eqnarray}
{\rm lim}_{f \to 1} \ \kappa (f) &&= \kappa_{\rm cr} <0 \ ,
\nonumber \\
{\rm lim}_{f \to 0} \ \kappa (f) && = 0 \ ,
\label{limkappa}
\end{eqnarray}
and $N_0=N$, $N_1=1$. 

Let us now consider the limit $N\to \infty$. If we take this limit at fixed small $f$ the physical curvature vanishes since $N_f \approx N$ in this regime and we obtain flat manifolds. The exception is the case $D=2$, for which the curvature is determined only by the gravitational coupling $f$. For higher dimensions, we can obtain an infinite graph of finite curvature only by letting the coupling $f$ become a function of $N$ such 
\begin{equation}
{\kappa(f(N))  N^{2/D} \over f(N) N_f} = \lambda \ ,
\label{renorm}
\end{equation}
where $\lambda$ is a fixed negative number independent of $N$. This means that the coupling $f$ becomes a running coupling whose $N$-dependence is set by the equation (we consider $N$ as a continuous variable for ease of presentation)
\begin{equation}
{d\over dN} {\kappa(f(N)) N^{2/D} \over f(N) N_f} =0 \ .
\label{running}
\end{equation}
Correspondingly, also the Planck length and the Newton constant become running couplings $\ell_P(N)$ and $G_N(N)$ defined via the dependence $f(N)$. Since $N_f \to N$ for $f \to 0$, the solution of (\ref{running}) in the limit $N\to \infty$ requires $f\to 0$. Of course this presupposes that $\kappa(f)$ vanishes slower than $f$ in this limit, which seems to be indeed the case for the investigated graphs. 
In higher dimensions, thus a careful limiting procedure is needed in which $f$ and $N$ are varied simultaneously. This is one of the reasons why higher-dimensional simulations are difficult with presently available computational power.  

The point to be retained as a conclusion of this section is that infinite curved graphs are discretizations of emergent manifolds of negative curvature, so-called Cartan-Hadamard manifolds \cite{shiga}, with curvature given by
\begin{equation}
\Lambda =  {\lambda \over \ell_p^2} \ .
\label{coco2}
\end{equation}
As we will show below, $-\Lambda$ can be interpreted as the cosmological constant. 

\section{Cartan-Hadamard manifolds in 2D}

In this section we will focus on the numerically accessible 2D case. We will thus be interested in infinite graphs at weak, but non-vanishing gravity. In this case most edges $(ij)$ support a number of squares $S_{ij}$ less than the maximal value $2D-2$ reached for a hypercubic lattice. Since $T_{ij}=0$, $P_{ij}=0$ (apart from rare defects), the Ollivier-Ricci curvature of the ground state, computed from (\ref{orc}) is negative. In Fig.2 and Fig.3 we present the average number of squares per vertex as a function of the coupling constant $g\hbar$. The maximum value for a hypercubic lattice is $2D(D-1)$. The quantum phase transition from locally tree-like random graphs to hypercubic lattices is clearly visible. 

Any finite graph has a genus, which is defined as the smallest genus of a surface on which it can be embedded as a 2-cell complex without any edge crossings \cite{genus}. Random regular graphs are expander graphs with high probability and expander graphs have a genus growing linearly with the number of vertices \cite{expander}. In 2D, the flat square lattice at vanishing gravity has genus 1. The random-to-geometric transition can thus be viewed also as a quantum transition in which the graph genus decreases. 

In order to ``geometrize" the graphs, we must turn these purely combinatorial objects into 1-skeletons of topological spaces by assigning a fixed length $\ell$ to the edges. When this is done, the 2-cell embeddings become 1-skeletons of tilings of Riemann surfaces. The combinatorial Ollivier curvature, $\propto (S_{ij}-(2D-2))/\ell^2 $ for each edge $(ij)$, has to be matched to a continuum Ricci curvature by choosing an appropriate metric on the surface. In general this is not easy. In particular cases, however it can be done, as we now show.

\begin{figure}[t!]
	\includegraphics[width=9.5cm]{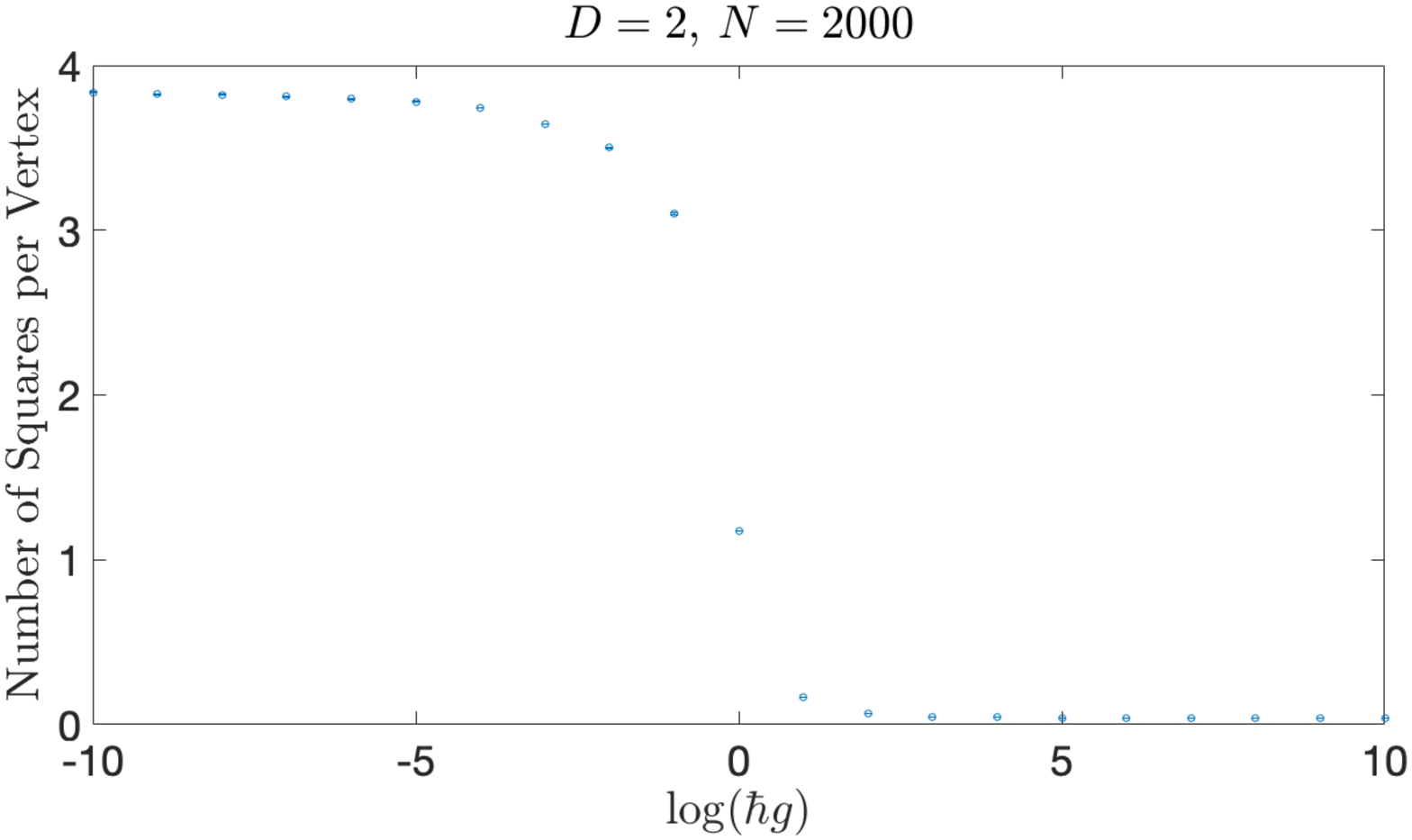}
	\vspace{-0.3cm}
	\caption{Average number of squares per vertex as a function of the coupling constant $g\hbar$ (normalized to 1 at the transition) for $D=2$. The maximum value for a square lattice is 4.}
	\label{Fig.2}
\end{figure}

\begin{figure}[t!]
	\includegraphics[width=9cm]{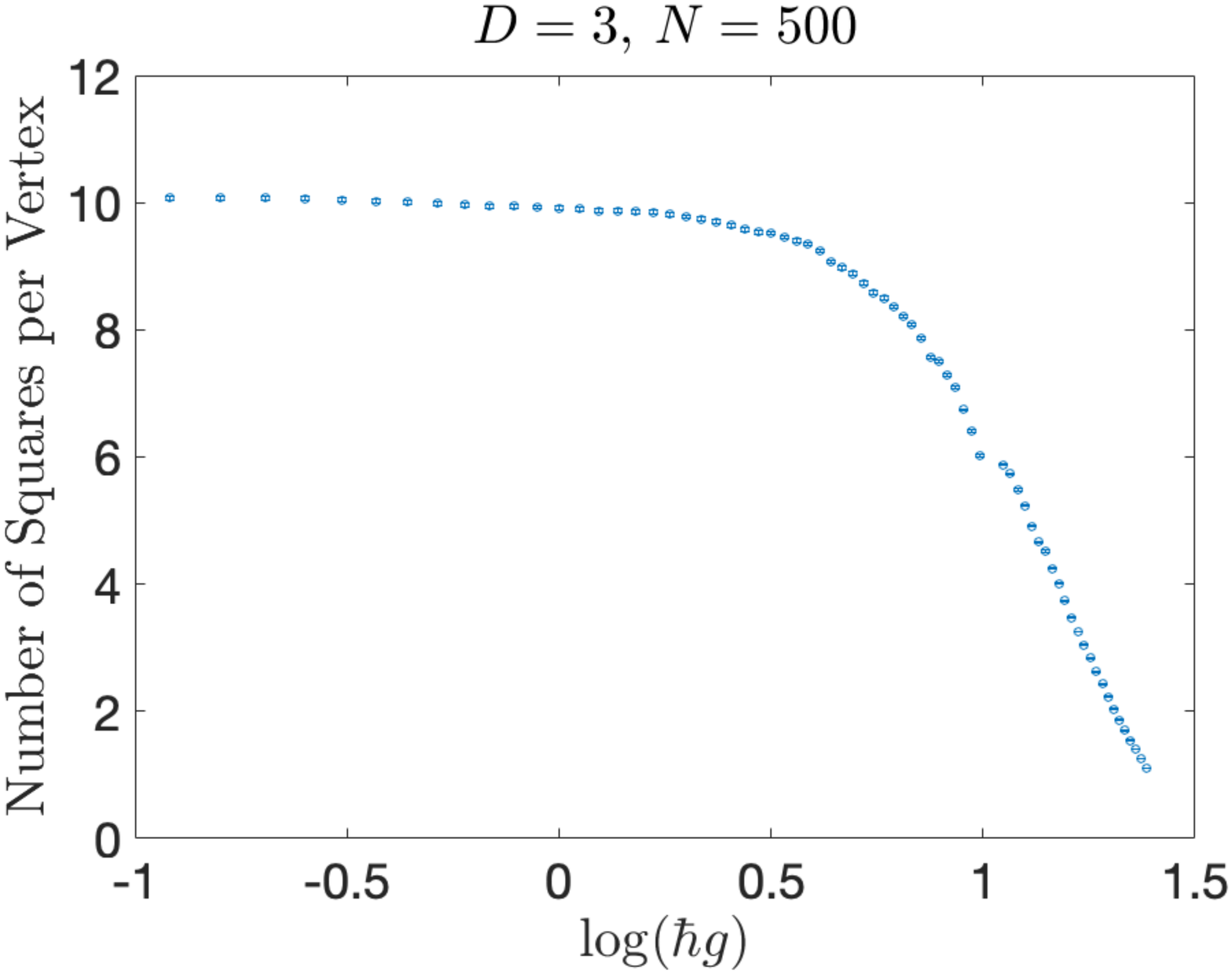}
	\vspace{-0.3cm}
	\caption{Average number of squares per vertex as a function of the coupling constant $g\hbar$ for $D=3$. The maximum value for a cubic lattice is 12.}
	\label{Fig.3}
\end{figure}

Let us focus on the 4-regular configuration with three squares per vertex on average, as shown in Fig.1. Actually, let us consider first a simpler example in which {\it each vertex} of the graph is common to {\it exactly} three squares and one other cycle of size larger than five (there are no pentagons in this phase). In this case, the number $F$ of faces i.e. cycles, is less than the maximum $N$ since more vertices are needed on average to form a cycle than in the pure square lattice. Therefore, the Euler characteristic $N-E+F = -N+F <0$ is negative and the genus of the graph is larger than 1, in accordance with its constant negative curvature. The infinite graph in the limit $N\to \infty$ is thus embedded in the universal cover of the genus $g>1$ surface relevant for finite graphs, i.e. a negative-curvature Cartan-Hadamard manifold, as derived in the generic case in the previous section. For unit negative curvature this manifold is the hyperbolic plane ${\mathbb H}^2$ and has the Poincar\'e disk representation (for a review see \cite{poincare}). In this case, any 4-regular graph with 3 squares and a cycle of length larger than 4 per vertex can be promoted to the 1-skeleton of a tiling of a negative curvature Cartan-Hadamard manifold by assigning a fixed length to each edge (CH) \cite{datta}. In Fig.4 we show the Poincar\'e disk representation of such a tiling for the case of the hyperbolic plane and the fourth cycle at a vertex being a hexagon. 

\begin{figure}[t!]
	\includegraphics[width=8cm]{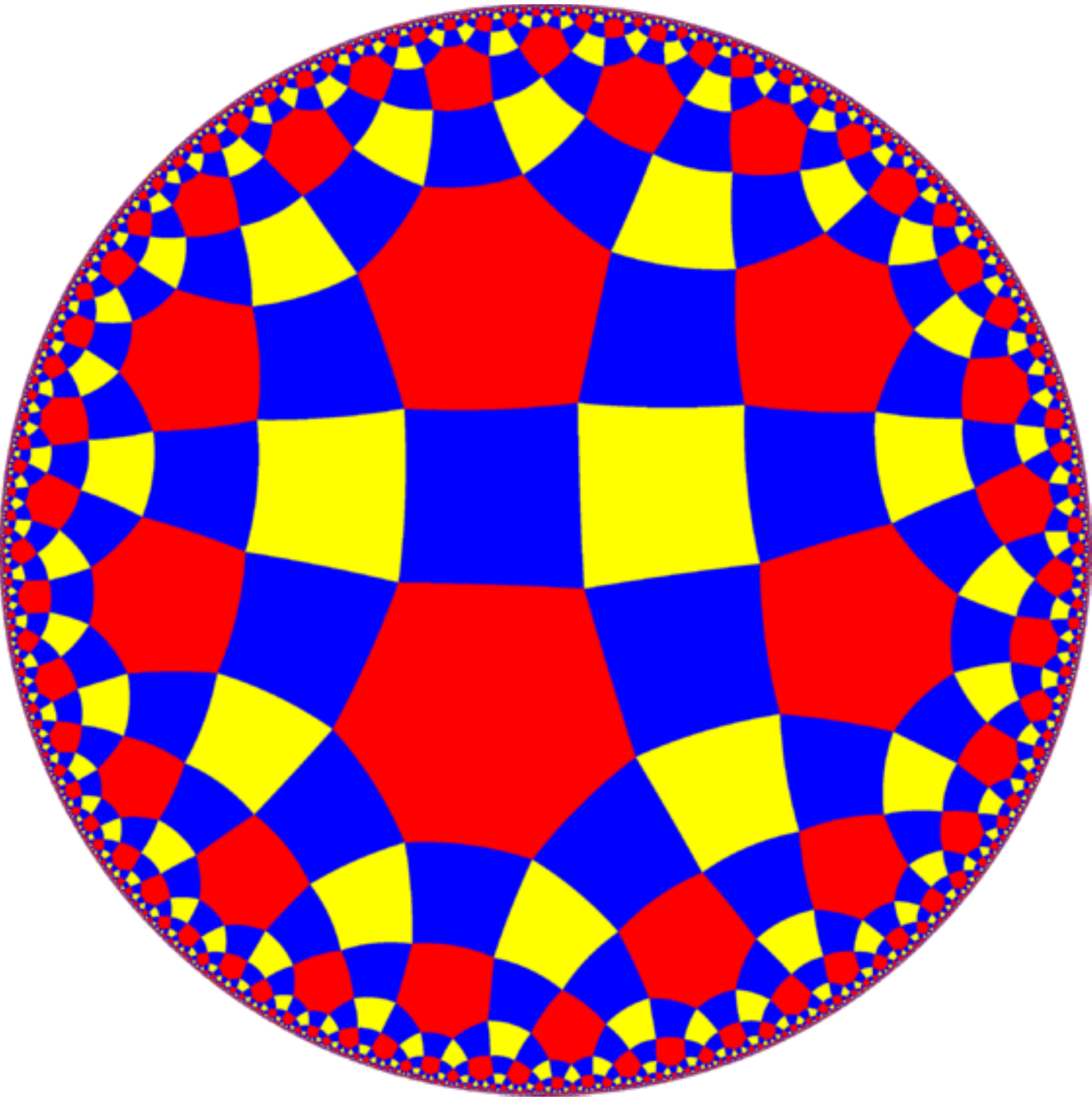}
	\vspace{-0.3cm}
	\caption{Tiling of hyperbolic space in the Poincar\'e disk representation with squares and hexagons. Each vertex is common to exactly three squares and one hexagon. }
	\label{Fig.4}
\end{figure}

In the general case, the number of squares per vertex is three only on average, the fourth cycles can be of any size larger than five and there can be triangle and pentagon defects, all of which are local variations upon the 1-skeleton in Fig.4. The negative curvature, however, implies that the genus is still larger than 1 and that the infinite graphs can be embedded in the Poincar\'e disk. As a consequence, ground-state infinite graphs are discretizations of negative-curvature CH manifolds \cite{shiga}. 

Further support for the emergence of CH manifolds in 2D is provided by the spectral data of the graphs. Let us consider the heat kernel $K(t)$, encoding the return probabilities after $t$ diffusion steps on the graph, 
\begin{equation}
K(t) = {\rm tr}\  {\rm e}^{\Delta t} \ ,
\label{return}
\end{equation}
where the (normalized) graph Laplacian $\Delta$ can be computed as
\begin{equation}
\Delta = -I +{1\over 2D} A \ ,
\label{glap}
\end{equation}
with $I$ the identity matrix and $A$ the graph adjacency matrix. The equivalent information is contained in the spectral function
\begin{equation}
d_{\rm s} (t) = -2 {d \ {\rm ln} K(t) \over d \ {\rm ln} t } \ .
\label{spectraldata}
\end{equation} 
The spectral function $d_s(t)$ probes the spectral dimension of the manifolds in different regimes, on microscopic scales for $t \to 0$, on large scales for $t \to \infty$. The spectral dimension is, roughly speaking, the number of independent directions available for a random walker. For an Euclidean manifold it coincides with its topological dimension and is independent of $t$, in general, however it can differ and needs not even be an integer for fractals and graphs. 

In Fig.5 we display the spectral function of the ground state graphs for $D=2$ and $N=2000$ as a function of the coupling constant $g\hbar$. One can readily identify several features. First of all, the phase transition from random graphs to ordered ones is clearly visible. In the ordered phase we can identify three regimes. For small values of the diffusion time $t$ we see a shape increasing to a maximum and then decreasing, exactly like in the random phase but on a much smaller scale. As discussed in the previous section, these are the discreteness effects below the Planck scale. At the largest diffusion times $t$ the curve $d_{\rm s}(t)$ decreases to zero: this is due to the finite-size effects. The most interesting regime for us is the intermediate regime between the minimum and the subsequent maximum. This is the region where we expect to probe the continuum properties of the emerging manifold. 

\begin{figure}[t!]
	\includegraphics[width=9cm]{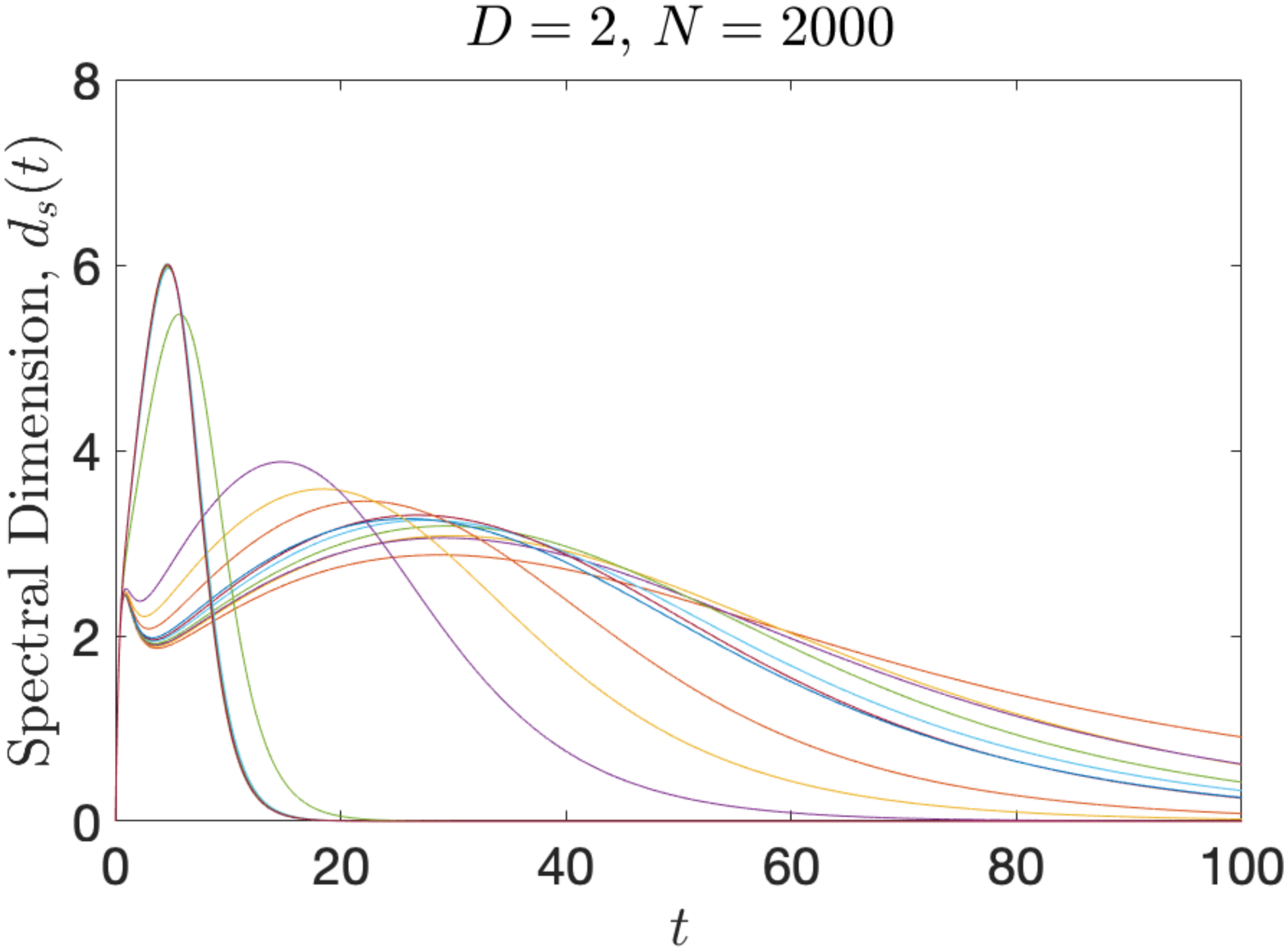}
	\vspace{-0.3cm}
	\caption{The spectral function $d_{\rm s}(t)$ of the 2D ground state graphs as a function of the coupling constant $g\hbar$. The phase transition is clearly identifiable in a change of behaviour as the coupling constant is lowered. The violet curve corresponds to the case of an average of three squares per vertex discussed above. }
	\label{Fig.5}
\end{figure}

An increasing $d_{\rm s}(t)$ in this regime is typical of constant negative curvature CH manifolds\cite{shiga}. The return probabilities on CH manifolds of constant negative curvature $(-H^2)$ have the uniformly continuous estimate \cite{davies, grigorian, dunne2} 
\begin{equation}
K(t) = {\left( 1+H^2 t \right)^{(D-3)/2} \over t^{D/2}} {\rm e}^{ -{(D-1)^2 \over 4} H^2 t} \ ,
\label{returnCH}
\end{equation}
which gives the spectral curve
\begin{equation}
d_{\rm s} (t)  = D- (D-3) \left( {H^2 t\over 1+H^2 t} \right) +{(D-1)^2\over 2} H^2 t \ .
\label{specCH}
\end{equation}.

In Fig.6 we show the fit of the intermediate regime of the violet spectral curve, corresponding to an average of three squares per vertex, to this functional form (a scale factor from the different definitions of the diffusion time $t$ can be absorbed in a redefinition of the curvature). The spectral dimension of the graph is obtained as $d_{\rm s} = 1.88 \pm 0.08$ and the dimension and curvature parameters have p-values of $1.1 \times \ 10^{-16}$ and $6.7 \times 10^{-6}$, respectively, confirming that they are both very relevant parameters. 
The flatter spectral curves in Fig.5 correspond to lower values of the coupling constant, for which the average number of squares per vertex is much closer to the maximum value 4 and, thus, the graphs are expected to have already genus 1. 

\begin{figure}[t!]
	\includegraphics[width=9cm]{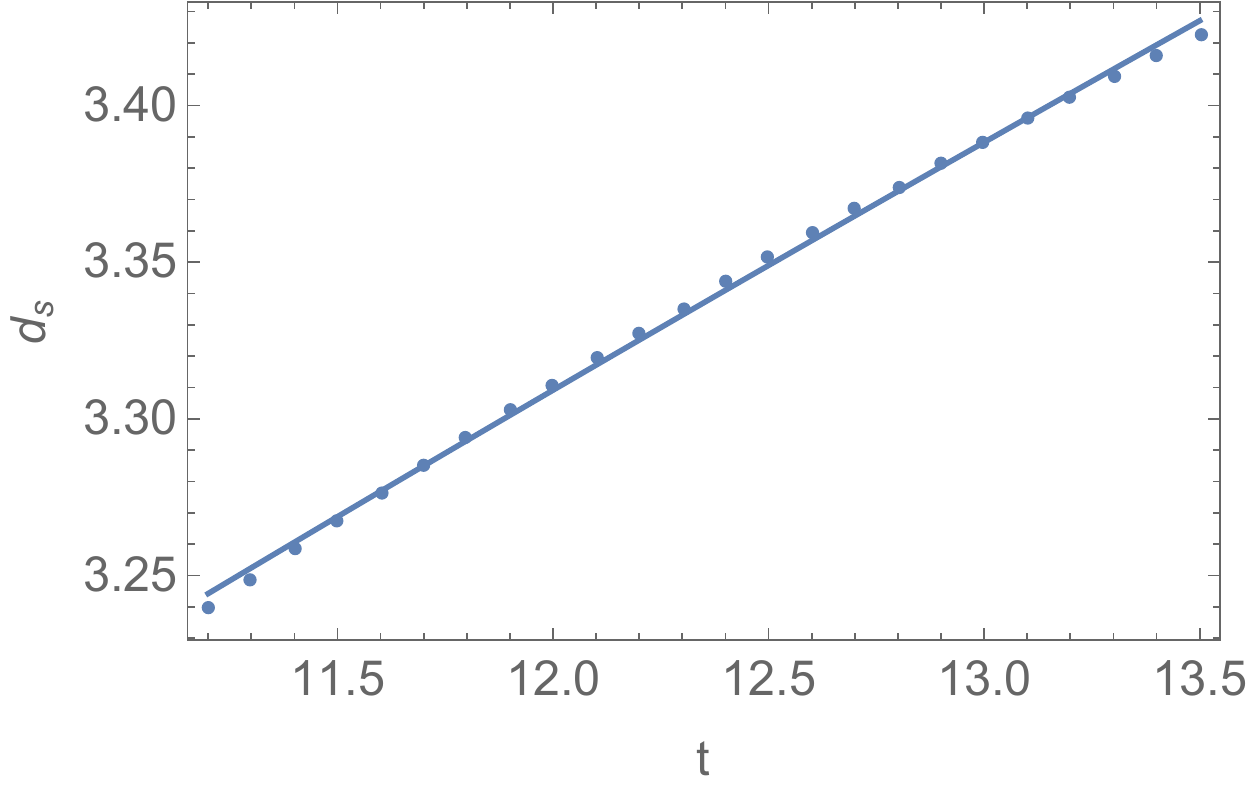}
	\vspace{-0.3cm}
	\caption{Fit of the intermediate regime of the ordered-phase violet curve in Fig.4, corresponding to an average of three squares per vertex, to the CH spectral function (\ref{specCH}).}
	\label{Fig.6}
\end{figure}

Unfortunately, the ``large-scale regime" is realized only in a narrow diffusion time window on such a small graph, squeezed between discreteness effects at small $t$ and finite size effects at large $t$. For comparison, in Fig. 7 we show the spectral function for a 3D,  $N=500$. graph. The quantum transition corresponding to the condensation of squares is clearly visible here too. However, the graph is much too small for extracting numerically the large scale behaviour. The computing power needed to address the large-scale behaviour in 3D is beyond computational resources presently available to the author. 

\begin{figure}[t!]
	\includegraphics[width=9cm]{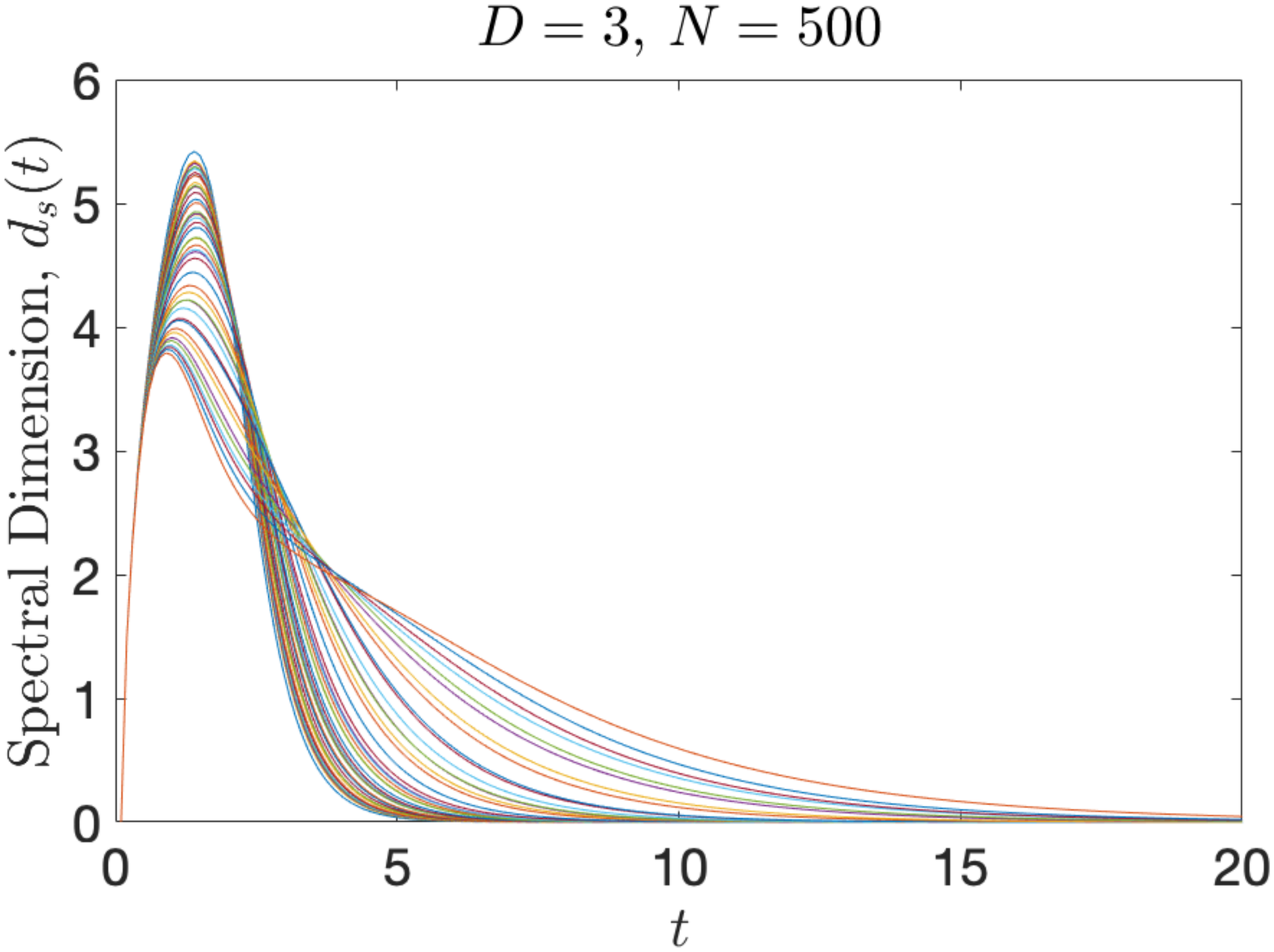}
	\vspace{-0.3cm}
	\caption{The spectral function $d_{\rm s}(t)$ of the 3D ground state graphs as a function of the coupling constant $g\hbar$. The phase transition is evident also here but graphs that can be treated with present computational power are way to small to identify the continuum regime in 3D. }
	\label{Fig.7}
\end{figure}

\section{Emergent time: from Cartan-Hadamard manifolds to de Sitter space-time} 

From now on we shall focus on the continuum properties of the emerging Cartan-Hadamard manifolds.
In particular, let us equip them with the only natural ``dynamics", Brownian motion \cite{ledrappier}. Brownian motion is characterized by the behaviour of the mean square displacement, or quadratic variation, $\langle \Delta x^2(t) \rangle$ (we focus on the 1D case for ease of presentation but everything is straightforwardly generalized to any number of dimensions). In the standard case we have the Einstein relation
\begin{equation}
\langle \Delta x^2(t) \rangle \propto t \ ,
\label{normaldiff}
\end{equation}
which is equivalent to the normal diffusion law
\begin{equation}
{\partial \over \partial t} u = D_{\rm diff} {\partial^2 \over \partial x^2} u \ ,
\label{normalheat}
\end{equation}
for the probability distribution $u$ of the diffusing particles, $D_{\rm diff}$ being the diffusion coefficient. Depending on the geometric properties of the underlying manifold, however, diffusion can be anomalous (for a review see \cite{dunne, sandev, procaccia}), with a characteristic walk dimension $d_w$ defined by
\begin{equation}
\langle \Delta x^2(t) \rangle \propto t^{2\over d_w} \ .
\label{anomalousdiff}
\end{equation}
A particularly interesting case is realized for anomalous diffusion with walk dimension $d_w=1$. In this case diffusion becomes {\it ballistic} and the probability distribution function $u$ is a weighted average of the initial data on the characteristic lines $(x+vt)=0$ and $(x-vt) =0$ of the hyperbolic operator $(\partial^2/\partial t^2 -v^2 \partial^2/\partial x^2)$ \cite{sandev, procaccia}. This means that the behaviour of the distribution function is governed by the wave equation
\begin{equation}
\left( {\partial ^2 \over \partial t^2} -v^2 {\partial ^2 \over \partial x^2} \right) u = 0 \ ,
\label{kleingordon}
\end{equation} 
where $v$ is the velocity of propagation of the ballistic particles (for a review see \cite{sandev, procaccia, chandrashekar}) and shows that ballistic diffusion is deterministic and equivalent to relativistic propagation on a Lorentzian manifold obtained by adding the diffusion time $t$ to the original space variables. The deterministic character of ballistic diffusion can also be recognized from the Taylor-Kubo formula for the autocorrelation function of velocities \cite{kubo},
\begin{equation}
\langle v(t+\tau) v(\tau) \rangle \propto t^{(2/d_{\rm w})-2} \ ,
\label{kubo}
\end{equation}
which becomes $t$-independent for $d_w=1$, indicating infinitely long memory. 

Diffusion is strongly influenced by geometry. It turns out that diffusion on a negative curvature Cartan-Hadamard (CH) manifold is ballistic, it escapes to infinity at unit speed and has an asymptotic direction \cite{prat, kendall, hsu} (for a review see \cite{hsubrief, hsubook, arnaudon}).  Not only it is ballistic, the diffusion time coincides asymptotically with one of the coordinates of the manifold itself. As we now show, the consequence is that diffusion on a negative curvature Riemannian manifold is asymptotically equivalent to the wave evolution of the probability distribution on a corresponding Lorentzian manifold of the same dimension and positive curvature.

To do so let us focus on the simplest case of two dimensions (2D), the generalization to any other number of dimensions is straightforward. The (radially symmetric) 2D Cartan-Hadamard manifold is the upper sheet of a two-sheeted hyperboloid, which we shall parametrize by the three-vector
\begin{equation}
{\bf x} = \begin{pmatrix} {1\over H} {\rm sinh} Hz \ {\rm cos} \theta \\ {1\over H} {\rm sinh} Hz \ {\rm sin} \theta \\ {1\over H} {\rm cosh} Hz \end{pmatrix} \ .
\label{twosheets}
\end{equation}
When embedded in Euclidean space this defines a manifold of positive curvature. However, when embedded in Minkowski space with metric $(+1, +1, -1)$ it becomes a 2D Riemannian manifold of negative constant curvature $(-H^2)$ and metric
\begin{eqnarray}
ds^2 &&= dz^2 + G^2(z) d\theta^2 \ ,
\nonumber \\
G(z) &&= {{\rm sinh} Hz\over H} \ .
\label{metrictwo}
\end{eqnarray} 
In the limiting case $H=1$ this is called the hyperboloid model for hyperbolic space. Note that the manifold is intrinsically Riemannian even if it is embedded in Minkowski space. 

Brownian motion on this manifold decomposes in two independent radial and angular Brownian motions $(z_t, \theta_t)$. The first one is called radial since $z$ determines the radius $(1/H) {\rm sinh} Hz$ of the manifold. Two aspects of these are paramount for our purposes \cite{prat, kendall, hsu} (for a review see \cite{hsubrief, hsubook, arnaudon}). First, 
\begin{equation}
\lim_{t\to \infty} \ z_t= {H\over 2} t \ ,
\label{rad}
\end{equation}
i.e. asymptotically, the coordinate $z$ can be identified with the diffusion time (note that a dimensionful diffusion coefficient has been absorbed into $t$) . Second, the angular Brownian motion converges to a limiting angle $\theta_\infty$ and its distribution function becomes uniform on the circle asymptotically at large times. In particular, the 
quadratic variation of the angular Brownian motion is given by \cite{hsubrief, arnaudon} 
\begin{equation} 
\langle \Delta \theta_t ^2 \rangle = \int_0^t ds {1\over G\left( z_s \right)^2} \ .
\label{quadvartot}
\end{equation} 

Let us consider this angular quadratic variation near the limiting angle $\theta_\infty$,
\begin{equation} 
\langle \left(\theta_\infty- \theta_t \right)^2  \rangle= \int_t^\infty ds {1\over G\left( z_s \right)^2} \ .
\label{quadvarinf}
\end{equation} 
If we consider large values of $t$ we can substitute $z_s$ with its limiting value (\ref{rad}). Then, using (\ref{metrictwo}) we obtain 
\begin{equation}
\langle \left(\theta_\infty-\theta_z \right)^2 \rangle = \tau^2 = 4 \ {\rm e}^{-2Hz} \ .
\label{limquadmot}
\end{equation}
Asymptotically, the angular Brownian motion is ballistic in terms of the new time scale $\tau = 2 \ {\rm exp}(-Hz)$ and the probability distribution becomes thus a function of the characteristic lines $\left(\theta_{\infty} -\theta \right)+ \tau =0$ and $\left( \theta_{\infty} -\theta \right)  -\tau=0$ of the hyperbolic equation
\begin{equation}
\left( {\partial^2 \over \partial z^2} + H {\partial \over \partial z} - 4H^2 {\rm e}^{-2Hz} {\partial^2 \over \partial \theta^2} \right) u(z, \theta) = 0 \ ,
\label{kleingordoncurved}
\end{equation}
whose general solution can be written as
\begin{equation}
u(z, \theta) = a_+ u_+ \left( \theta -\theta_\infty + 2 {\rm e}^{-Hz} \right) + a_- u_- \left( \theta -\theta_\infty - 2 {\rm e}^{-Hz} \right) \ .
\label{general}
\end{equation}

This hyperbolic operator defines the wave equation on a Lorentzian curved manifold. To see this, let us consider the one-sheeted hyperboloid parametrized by
\begin{equation}
{\bf x} = \begin{pmatrix} {1\over H} {\rm cosh} Hz \ {\rm cos} \theta \\ {1\over H} {\rm cosh} Hz \ {\rm sin} \theta \\ {1\over H} {\rm sinh} Hz \end{pmatrix} \ .
\label{onesheet}
\end{equation}
When embedded in Euclidean space this is a manifold of negative curvature. When embedded in Minkowski space with metric $(+1, +1, -1)$, however it becomes a Lorentzian manifold of constant positive curvature $H^2$ and metric
\begin{equation}
ds^2 = -dz^2 + {{\rm cosh^2}Hz\over H^2} d\theta^2 \ .
\label{metricone}
\end{equation}
This manifold is known as de Sitter space-time (for a review see \cite{strominger, cosconst}). De Sitter space-time is the model for an empty universe with a positive cosmological constant $H^2$ and an inflationary (exponential) expansion and is considered as a realistic model for the universe in the far future, when the cosmological constant comes to dominate the matter density, diluted by expansion. For large values of $z$, the upper sheet of the two-sheeted hyperboloid and the one-sheeted hyperboloid coincide up to exponentially small corrections since both approach the limiting cone. Networks on de Sitter space-time were investigated in \cite{krioukovdesitter}.

For values of $z \gg 1/H$, the wave equation for a probability distribution $u(z, \theta)$ on de Sitter space-time becomes exactly (\ref{kleingordoncurved}). After diffusion time $t$ gets soldered to one manifold coordinate by (\ref{rad}), ballistic diffusion on a negative-curvature Riemannian manifold becomes equivalent to the wave equation for the probability distribution on a corresponding Lorentzian manifold of positive curvature. Therefore, in this picture, time as a coordinate of a Lorentzian manifold is an emergent concept which becomes valid only asymptotically for $z\gg 1/H$. Since the isometry group of the manifold is $SO_{+}(D, 1)$ the emergent time comes automatically with an arrow. 

The picture just described concerns the closed slicing of de Sitter space, with space a positive curvature closed manifold. Let us consider, instead the flat slicing \cite{strominger, cosconst}. In this case the negative-curvature hyperboloid sheet has the parametrization
\begin{equation}
{\bf x} = \begin{pmatrix} {x\over H} {\rm exp} (Hu) \\ {1\over H} {\rm sinh} Hu -{1\over 2H} x^2 {\rm exp}(Hu) \\ 
{1\over H} {\rm cosh} Hu +{1\over 2H} x^2 {\rm exp}(Hu)
\end{pmatrix} \ ,
\label{twoflat}
\end{equation}
with $u \ge 0$ and metric given by 
\begin{equation}
ds^2 = du^2 + {{\rm e}^{2Hu}\over H^2} dx^2 \ ,
\label{metriflattwo}
\end{equation}
while de-Sitter space-time with positive curvature is parametrized by
\begin{equation}
{\bf x} = \begin{pmatrix} {x\over H} {\rm exp} (Hu) \\ {1\over H} {\rm cosh} Hu -{1\over 2H} x^2 {\rm exp}(Hu) \\ 
{1\over H} {\rm sinh} Hu +{1\over 2H} x^2 {\rm exp}(Hu)
\end{pmatrix} \ ,
\label{oneflat}
\end{equation}
with $-\infty < u < +\infty$ and the inflationary metric
\begin{equation}
ds^2 = -du^2 + {{\rm e}^{2Hu}\over H^2} dx^2 \ .
\label{metricflatone}
\end{equation}
The two times $z$ and $u$ coincide asymptotically for large values. Therefore the same picture as above is valid in flat slicing. For $u\gg 1/H$ the two manifolds coincide up to exponentially small corrections and the Lorentzian picture with $u$ as coordinate time is a valid alternative description of ballistic diffusion on the negative-curvature manifold. However, this alternative description cannot be extrapolated back for small values of $u$. In particular, the big bang at $u=-\infty$ is avoided, for $u<1/H$ the universe becomes a flat Euclidean manifold, as in the Hartle-Hawking scenario \cite{hh, hhrev}. 

To conclude this section, note that, given that the metric is $z$-dependent, Brownian motion can also be considered as a random process for the metric itself. In this case, the asymptotic behaviour of this process coincides with the usual exponential expansion of the de Sitter universe.

\section{Boundary and dimension at infinity}

Every geodesically complete, D-dimensional manifold $M$ of constant negative curvature has a geometric boundary manifold $\partial M$ defined as the locus of all equivalence classes of geodesic rays that remain in bounded distance of each other. If $M$ is simply connected, the boundary $\partial M$ is a sphere $S^{D-1}$ (see e.e. \cite{ratcliffe}). If we adopt the Lorentzian point of view, the boundary $\partial M$ can be viewed as the boundary of space-time. There is a one-to-one correspondence between interior $SO_{+}(D,1)$ isometries on $M$ and (D-1)-dimensional conformal transformations on the boundary $\partial M$. This correspondence is, of course, a Riemannian equivalent of the holographic principle on Lorentzian negative-curvature manifolds like anti-de-Sitter (adS) space \cite{thooft, susskind} (for a review see \cite{bousso}). Via the emerging time coordinate induced by ballistic motion of test particles, however, it becomes asymptotically a de-Sitter version of holography. Using, e.g. flat slicing, as discussed in the previous section, we obtain an asymptotic one-to-one correspondence between conformal field theories on the boundary and bulk theories with the local de Sitter special relativity group $SO_{+}(D,1)$ \cite{dSrel1} (for a review see \cite{dSrel2}), with the added bonus of an emerging arrow of time. There is, however, a difference with respect to the usual framework. While the topological dimension of the boundary is $(D-1)$ as in the familiar case, its spectral dimension is always 3, implying a 2D spatial boundary, independent of the bulk topological dimension, as we now discuss. This is of course compatible with any overall topological dimension of the universe, the two dimensions being different concepts.

In general, diffusion processes probe the intrinsic geometry of a manifold. One would expect that, in the limit $t\to \infty$ of infinite diffusion time, the diffusion process on $M$ probes the ``geometry at infinity" of the boundary $\partial M$. However, this is not so, because the Laplacian on a constant negative curvature manifold has a spectral gap
\begin{equation}
\lambda_0 = - {\rm lim}_{t\to \infty} {{\rm ln}\ K(t) \over t} = {(D-1)^2\over 4}  \ ,
\label{spegap}
\end{equation} 
representing the bottom of the spectrum of the positive operator $-\Delta$, with $K(t)$ the return probability kernel (\ref{returnCH}). As a consequence of the spectral gap, the return probabilities $K(t)$ are dominated by an exponential behaviour at large $t$ and the boundary at infinity is hidden for any fixed observer that monitors the diffusion process. In other words, because of ballistic diffusion, this process runs away to infinity extremely quickly, in a fixed asymptotic direction and never comes back for large times. The spectral gap, due to negative curvature, is responsible for both the presence of the boundary and for ballistic diffusion. 

If one wants to probe the geometry at large scales one must define a modified, slower diffusion process, as seen by a ``co-diffusing" observer, so as to ``subtract" the spectral gap and keep only the next-higher eigenvalues of the Laplacian. This can be done and this slower diffusion process is called the infinite Brownian loop \cite{anker}. Let us denote by $\varphi_0$ the most symmetric eigenstate of the Laplacian corresponding to the lowest eigenvalue $\lambda_0$ (it can be expressed in terms of a Jacobi function \cite{jacobi}), $\left( \Delta + \lambda_0 \right) \varphi_0 = 0$. The co-diffusing process is then the relativized $\varphi_0$-process \cite{sullivan}, with generator
\begin{equation}
\tilde \Delta (f) = {1\over \varphi_0} \Delta \left( f\varphi_0 \right) + \lambda_0 f = \tilde \Delta f + 2 \nabla {\rm ln} \ \varphi_0 \cdot \nabla f \ .
\label{relproc}
\end{equation}
As anticipated, in this process measured relatively to the ground state $\varphi_0$, the spectral gap falls out. This corresponds to dropping the exponential in (\ref{returnCH}), which is an expression of the central local limit theorem \cite{cllt}. The corresponding modified spectral function is 
\begin{equation} 
\tilde d_{\rm s} (t)  = D- (D-3) \left( {H^2 t\over 1+H^2 t} \right) \ ,
\label{modspe}
\end{equation}
whose infinite-time limit is 
\begin{equation}
D_{\rm inf} = {\rm lim}_{t \to \infty} \ \tilde   d_s(t) = 3 \ ,
\label{dimthree}
\end{equation}
independently of $D$. The quantity $D_{\rm inf}$ is called the ``pseudo-dimension", or ``dimension at infinity" of a constant negative curvature manifold \cite{anker}. It measures the spectral dimension of the manifold on large scales, for which $H^2 t \gg1$, as probed by the slow random process in the radial direction (co-diffusing process). This slow random process always sees three Euclidean dimensions and it is confined to the Weyl chamber \cite{anker}, which means, in this case, that it is confined to the forward direction, another expression of the emergent arrow of time.

\section{Conclusion}
We have constructed the following picture. General relativity, with time as a coordinate of a Lorentzian space-time manifold with positive cosmological constant becomes a correct description of the universe only asymptotically. This model cannot be extrapolated backwards, at ``earlier times" the fundamental negative curvature Riemannian nature of the universe dominates and dynamics is ballistic diffusion: the universe looks like a shuttlecock and there is no more time coordinate as we know it, only diffusion time. This picture is reminiscent of the Hartle-Hawking universe \cite{hh} (for a recent review see \cite{hhrev}), although it was derived from completely different principles. 

The universe has a boundary manifold and spatial spectral dimension three independently of the interior topological dimension. There is a one-to-one correspondence between interior isometries and conformal transformations on the boundary, an asymptotic de-Sitter version of the holographic principle.

As mentioned before, much larger and detailed numerical simulations are required for a full confirmation of the emergence of CH manifolds from graphs in higher dimensions.

\begin{acknowledgments}
We thank Christy Kelly, Fabio Biancalana, Gerald Dunne J.-H. Eschenburg and F. Ledrappier for discussions. 
\end{acknowledgments}

\end{document}